\documentclass[sigconf,authorversion,nonacm]{acmart}

\usepackage{algorithm}
\usepackage{color}

\usepackage{amsmath,amsfonts}
\usepackage{algorithmic}
\usepackage{graphicx}
\usepackage{textcomp}
\usepackage{xcolor}
\usepackage{booktabs}
\usepackage{tikz}
\usepackage{multirow}
\usepackage{booktabs}
\usepackage{comment}
\usepackage{subfigure}
\usepackage{amsmath}
\usepackage{xspace}
\usepackage{balance}
\usepackage{todonotes}
\usepackage{caption} 
\usepackage{newtxmath}
\usepackage{dblfloatfix}
\usepackage{gensymb}
\usepackage{lipsum} 
\usepackage[]{mdframed}
\usepackage{tcolorbox}

\newcommand{\system}{\emph{HE-Diffusion}\xspace}

\begin{document}


\title{Privacy-Preserving Diffusion Model Using Homomorphic Encryption}


\author{Yaojian Chen}
\affiliation{
\institution{SEIT Lab}
  \institution{Michigan State University}
  \city{East Lansing}
  \state{MI}
  \country{USA}
  \postcode{48824}
}

\author{Qiben Yan}
\affiliation{
\institution{SEIT Lab}
  \institution{Michigan State University}
  \city{East Lansing}
  \state{MI}
  \country{USA}
  \postcode{48824}
}

\author{Lin Gan}
\affiliation{
  \institution{Tsinghua University}
  \city{Beijing}
  \country{China}
}

\author{Guangwen Yang}
\affiliation{
  \institution{Tsinghua University}
  \city{Beijing}
  \country{China}
}


\begin{abstract}
In this paper, we introduce a privacy-preserving stable diffusion framework leveraging homomorphic encryption, called \system, which primarily focuses on protecting the denoising phase of the diffusion process.  \system is a tailored encryption framework specifically designed to align with the unique architecture of stable diffusion, ensuring both privacy and functionality.  To address the inherent computational challenges, we propose a novel min-distortion method that enables efficient partial image encryption, significantly reducing the overhead without compromising the model's output quality.  Furthermore, we adopt a sparse tensor representation to expedite computational operations, enhancing the overall efficiency of the privacy-preserving diffusion process. We successfully implement HE-based privacy-preserving stable diffusion inference. The experimental results show that \system achieves 500 times speedup compared with the baseline method, and reduces time cost of the homomorphically encrypted inference to the minute level. Both the performance and accuracy of the \system are on par with the plaintext counterpart. Our approach marks a significant step towards integrating advanced cryptographic techniques with state-of-the-art generative models, paving the way for privacy-preserving and efficient image generation in critical applications. \textbf{Source code is available at \url{https://github.com/HE-diffusion/HE-diffusion/}}. 
\end{abstract}

\maketitle

\section{Introduction}

The advent of stable diffusion models has marked a significant milestone in the field of generative artificial intelligence (GenAI), enabling the synthesis of high-fidelity images from textual descriptions. These models, exemplified by their capacity to transform abstract prompts into detailed visual content, have opened new avenues in digital content creation, personalized media, and data augmentation. However, the computational intensity inherent in the inference phase of stable diffusion models necessitates the offloading of these tasks to powerful server-side infrastructure, thereby introducing significant privacy concerns. The crux of the issue lies in the exposure of sensitive information to the server, including user-generated prompts and the resultant images, which may contain or infer personal or proprietary data. 

The benefits of implementing privacy-preserving mechanisms are manifold. It can enhance user trust, allow multiple parties to contribute private data or participate in the generative processes, and enable the use of stable diffusion models in fields where data sensitivity has traditionally been a barrier, such as healthcare, personal security, and confidential design. 

In prior research, CryptoNets~\cite{GiladBachrach2016CryptoNetsAN} demonstrated the use of neural networks on encrypted data, laying the groundwork for subsequent studies in this critical area. With the rapid development of GenAI, the intersection of deep learning, image processing, and cryptography has garnered substantial interest from researchers. Previous studies~\cite{ConcreteML, chen2022x} have investigated the security and privacy of GenAI models. Specifically, with regard to the stable diffusion model, existing work~\cite{dockhorn2022differentially, carlini2023extracting} identified the security issues, but they mainly focused on the security and privacy of training phase rather than on inference.

\emph{Homomorphic Encryption (HE)}~\cite{Gentry2009FullyHE} has become a popular method to conduct full computation on the encrypted data, which can alleviate users' privacy concerns during the computing escrow service. However, to achieve practical homomorphically encrypted stable diffusion inference is a challenging task, mainly due to the following two fundamental limitations of HE mechanisms. \emph{First}, HE only supports polynomial computation, and \emph{second}, it incurs high computation overhead (i.e., $10^4-10^5$ times of its plain-text counterparts)~\cite{samardzic2021f1} that could negatively impact its practical usage. Complicated structure of stable diffusion model seems to signal that fully encrypted inference is infeasible. 
However, this paper refutes this viewpoint. 
The objective of this paper is to architect such a HE-based framework that, for the first time, facilitates secure inference for stable diffusion models, providing an elevated level of security for most  users. 

In this paper, we introduce a novel HE-based framework, \system, to achieve practical privacy-preserving diffusion model inference. We carefully scrutinize nonlinear functions, strategically avoiding most of them to bypass the need for approximation to maintain accuracy. Considering the significant computational overhead of HE, we design a partial encryption scheme to drastically reduce the computational costs. Our methodology extends the line of research by introducing a privacy-preserving stable diffusion model, which, to our knowledge, is the first to apply HE  to the denoising phase of the diffusion process, thereby enabling secure generative modeling of images. By bridging the gap between advanced cryptographic techniques and state-of-the-art generative models, our work paves the way for privacy-preserving generative tasks. This private stable diffusion framework not only broadens the applicability of generative models in privacy-sensitive domains, but it also sets a precedent for the integration of security considerations in the burgeoning field of GenAI. Our project website with \system source code can be accessed through \url{https://github.com/HE-diffusion/HE-diffusion/}.

In summary, this paper makes the following contributions.
\begin{itemize}
\item We design \system, an efficient adaptation of the stable diffusion process that accommodates the constraints of HE. \system focuses particularly on protecting the denoising phase, which is essential for generating coherent images. 

\item We develop a series of optimization strategies, including a min-distortion method
for partial image encryption
and the implementation of sparse tensor representations, to enhance the computational efficiency of the encrypted inference process.

\item We have conducted extensive experiments to demonstrate that \system achieves 500 times speed-up for the HE denoising. This advancement brings the runtime of \system and plaintext stable diffusion to the same order of magnitude, resulting in a negligible loss in accuracy.
\end{itemize}

\section{Background and Related Work}

In the domain of secure computation and privacy-preserving technologies, our work intersects with several key areas, including HE, secure image processing, and the application of cryptographic techniques to deep learning models. Here, we contextualize our contributions within the landscape of existing research, particularly focusing on advancements in efficient HE schemes and secure image manipulation.

\subsection{Private Diffusion Model}
Among the generative models, stable diffusion models~\cite{latentdiffusion} have garnered significant attention for their capacity to generate high-quality, diverse samples, outperforming traditional generative adversarial networks (GANs)~\cite{goodfellow2020generative, park2019gaugan} in various benchmarks. The core principle of diffusion models involves gradually denoising a signal, typically starting from random noise, to generate coherent and complex outputs. While the potential of diffusion models in generating photorealistic images and their applications in content creation is undisputed, two critical concerns emerge when deploying these models in real-world scenarios: (1) the computational overhead associated with their inference process and (2) the privacy concerns associated with generating data from sensitive or personal information.

\begin{figure*}[htbp]
\centerline{\includegraphics[width=0.85\textwidth]{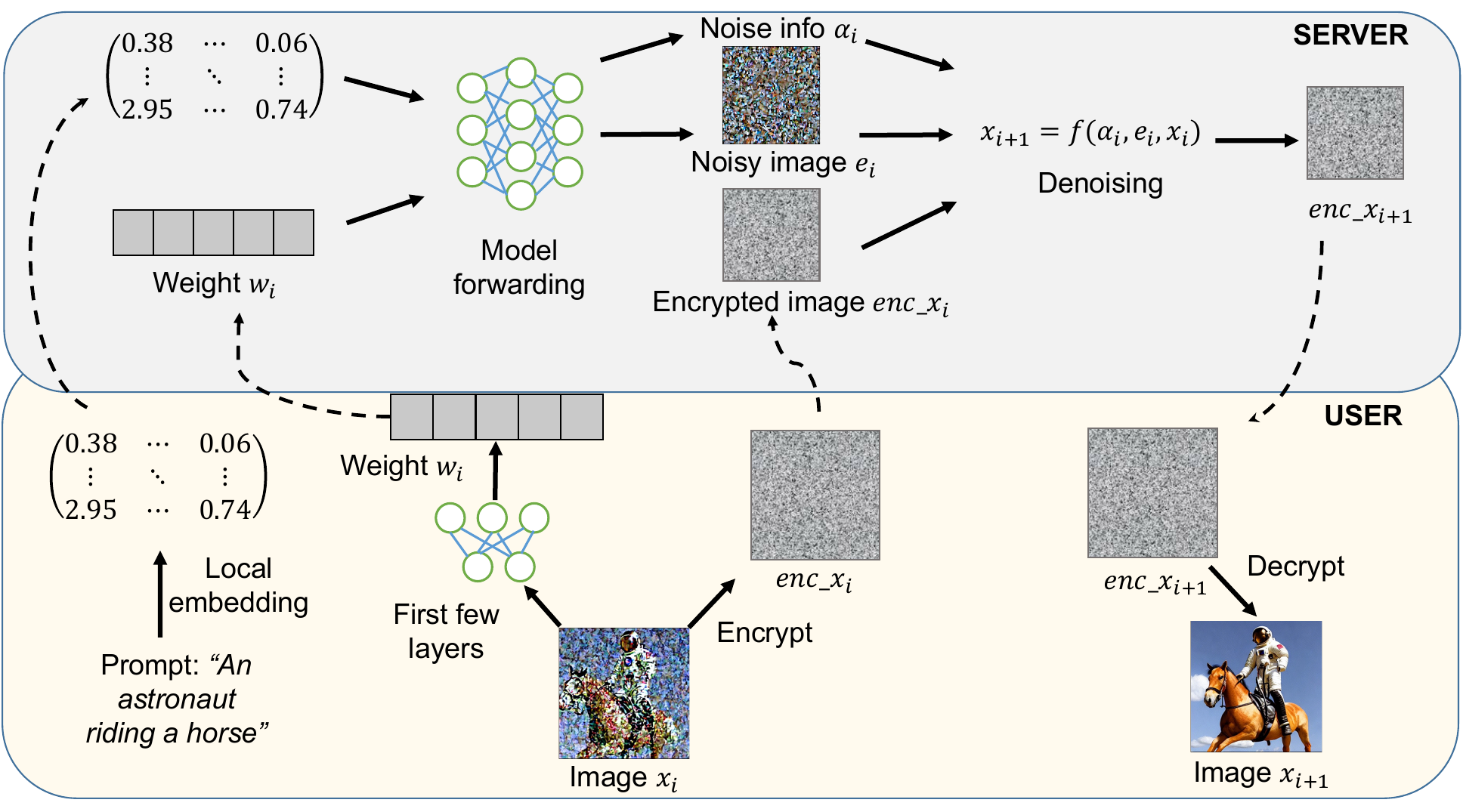}}
\caption{\textbf{Overview of \system at iteration $i$. Users provide embedded prompts tensors, intermediate weights, and the encrypted image $enc\_x_i$. Prompts are embedded locally by users, and the tensor is uploaded to server. The first few layers of the neural network are executed by users. The server performs forward propagation and denoising operations. After that, the server returns the encrypted result, i.e. $enc\_x_{i+1}$, to users.}}
\label{diffusion}
\end{figure*}

Although operating in a latent space greatly reduces the computation cost, traversing an inference process still requires powerful computational resources. Figure~\ref{diffusion} shows how the sampling works in one iteration. Each step involves a forward pass through a neural network.  Generating a single image can require hundreds to thousands of these passes, leading to significant computational overhead. Recent efforts aim to mitigate this overhead through various means, including model optimization and the introduction of more efficient sampling techniques~\cite{song2020score}, yet the computational demands remain a bottleneck for widespread adoption. To enable a broad spectrum of users to experiment with stable diffusion models, even those with limited computational resources, cloud computing and hosted services have gained popularity. However, the potential untrustworthiness of cloud servers and hosts introduces significant security and privacy concerns.

This work aims to develop privacy-preserving stable diffusion inference as a service via HE. The majority of the computation will be conducted on the server, while sensitive data remains in the possession of users. Previous works on private diffusion model mainly focused on protecting the training data~\cite{carlini2023extracting, hu2022privacy}, with emphasis on differential privacy~\cite{dockhorn2022differentially} and protection against membership inference attacks~\cite{matsumoto2023membership}. Protecting training data is crucial for model creators, whereas safeguarding the inference process is essential for the majority of users to maintain their privacy. However, discussion on inference privacy has been notably scarce. In fact, privacy concerns intensify when the inference process involves personal or sensitive data. In other deep learning applications, protecting users' private input and output is regarded as a significant concern~\cite{schwartz2011pii, boemer2020mp2ml}. Techniques such as HE have been applied to keep the sensitive information secret~\cite{jang2022privacy}. Considering that text-to-image and image refinement have become increasingly popular services, the privacy protection of stable diffusion inference is especially noteworthy. 

\subsection{Homomorphic Encryption}
The intersection of HE and generative models presents a promising avenue for safeguarding user privacy. HE enables the computation on ciphertexts, which has been utilized in various generative models. HE-transformer and its extensions~\cite{boemer2019ngraph,boemer2019ngraph2,boemer2020mp2ml, chen2022x} introduce private inference for transformer-based models. 
 Concrete-ML~\cite{ConcreteML} leverages large language models (LLMs) such as GPT-2~\cite{lagler2013gpt2} for privacy protection. It is evident that the research trend is moving towards providing privacy-preserving generative inference.

The practical implementation of HE has seen significant advancements through the development of specialized libraries and frameworks, designed to simplify the integration of HE into various applications. SEAL~\cite{sealcrypto} provides a comprehensive suite of functionalities for both fully and partially HE schemes, especially the Brakerski-Gentry-Vaikuntanathan (BGV) scheme ~\cite{Brakerski2014FullyHE} and the Cheon-Kim-Kim-Song (CKKS) scheme~\cite{Cheon2017HomomorphicEF} for approximate arithmetic. They have greatly improved the efficiency and practicality of HE. TenSEAL~\cite{tenseal2021} offers a seamless integration with PyTorch and emerges as a bridge between HE and machine learning. Concrete-ML~\cite{ConcreteML} abstracts the complexities associated with HE, offering a pathway for machine learning models to operate on encrypted data with minimal modifications. These developments underscore a pivotal shift towards making HE more accessible and practical for a broad spectrum of AL/ML applications. 

However, current attempts fall short in addressing HE's key challenge -- the computation overhead. A recent study~\cite{chen2018logistic} shows that homomorphically encrypted logistic regression is more than ten thousands of times slower than the plaintext version, using SEAL library~\cite{sealcrypto}. Moreover, most of relevant HE libraries lack of support for GPUs or other accelerators, which limits their efficient utilization of computing resources. The existing GPU-accelerated HE schemes lack sufficient integration to be usable~\cite{jung2021over}. Special accelerators~\cite{samardzic2021f1} may counteract HE’s overheads and enable wider applications. However, they require redesigning hardware architectures, making them scarcely accessible. There are some previous studies~\cite{de2021does, cao2014high, wang2013exploring} aiming to introduce techniques in high performance computing to enhance the efficiency of HE. Yonetani et al.~\cite{yonetani2017privacy} utilized sparsity to reduce encryption time. High performance algorithms~\cite{reagen2021cheetah} can achieve significant speedup, even with limited computational power. This is particularly relevant to our work's application scenario, where users cannot determine if the server possesses special hardware tailored for specific applications. With the generic computing resources, algorithm-level optimization demonstrates its benefits in terms of universality and cost-effectiveness.


\begin{figure}[htbp]
\centerline{\includegraphics[width=0.50\textwidth, height=0.13\textwidth]{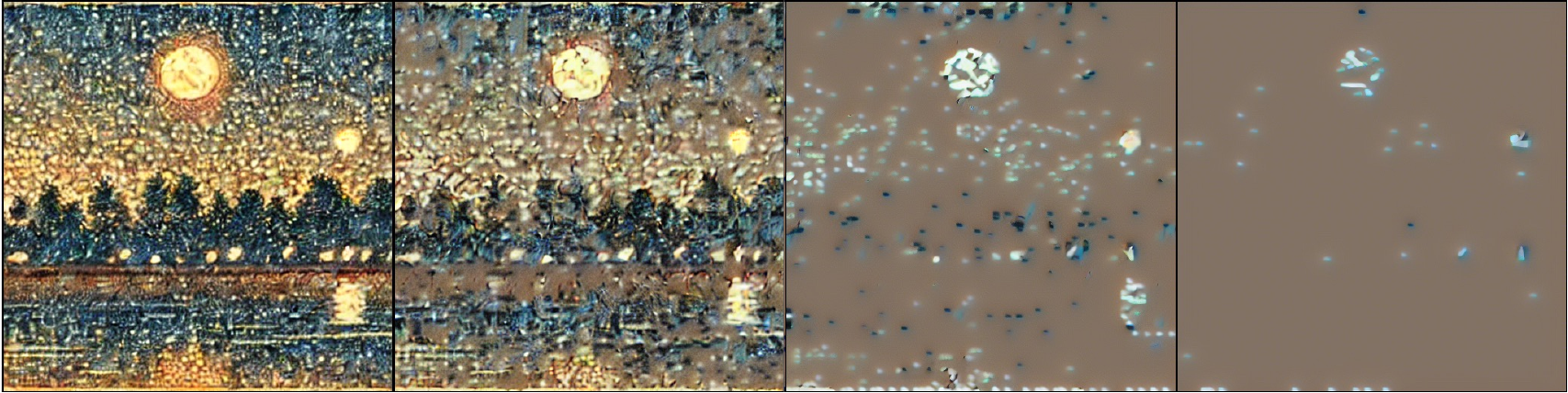}}
\caption{\textbf{When adding distortion in the latent space, the image changes as the figure shows. Specifically, we randomly set elements to 0 in the latent space to generate images with different distortion from the original image. Then, we transform them back to the pixel space. As the distortion grows, from top left to bottom right, the images become noisier and preserve less information.}}
\label{distortion}
\end{figure}

On the other hand, while the integration of generative AI with HE is promising, it poses a critical problem with accuracy loss. Neural networks introduce various non-linear activation functions and normalizations. However, under HE, these functions must be approximated because HE supports only addition and multiplication operations~\cite{Gentry2009FullyHE} (i.e., polynomial functions). Previous research for HE-inference~\cite{chen2022x, ao2023autofhe} has made substantial progress in achieving higher precision in these approximations. Other studies~\cite{jha2021deepreduce, peng2023autorep} aim to reduce the number of nonlinear operations. Nevertheless, errors still accumulate throughout the process due to the extensive use of nonlinear operations.

\subsection{Distortion and Information Loss}
Distortion refers to any alteration or degradation in the carrier medium (such as an image, video, or audio file) resulting from the embedding of hidden information. It is primarily applied in steganography~\cite{li2011survey} to hide the embedding information~\cite{fridrich2007practical, agrawal2021minimum}. Since distortion measures the distance between the original and modified images, it serves as a quantifier for information loss after data removal. If there is a low distortion between two images, it implies that the primary information is retained, and only unimportant sections without key semantics have been discarded. Another advantage of distortion is its applicability in various linear spaces, surpassing the capabilities of pixel-based methods. This implies its functionality in the latent space~\cite{rombach2022high}, where most calculations occur. 

Figure~\ref{distortion} validates the idea. When the elements are set to 0, we artificially discard information. Regarding the original image as an embedding of the revised one, distortion quantifies the informational difference. As a result, we can use distortion to quantify information loss.

In practice, additive distortion~\cite{filler2010gibbs} is often used due to its salient nature, which is defined as~\cite{fridrich2007practical}:

\begin{equation}
    D(X, Y) = \sum_{i,j} \rho_{ij} |x_{ij} - y_{ij}|,
\end{equation}
where $X$ and $Y$ are the original image and the modified image, $\rho$ is the cost matrix. The cost matrix $\rho$, a central parameter in distortion calculation, depends only on the original image $X$. There are different designs of the cost functions, known as HUGO~\cite{pevny2010using}, WOW~\cite{holub2012designing}, HILL~\cite{li2014new}, 2D-SSA based method~\cite{xie2021new} and QMP~\cite{liu2024efficient}. These methods aim to make $X$ and $Y$ resemble each other to counter steganalysis. Since the definition of the cost function is independent from the form of additive distortion, we can choose any effective cost function for convenience of implementation.

\section{Threat Model}
In this paper, the attacker's goal is to compromise the confidentiality of data processed by the diffusion model. Specifically, the adversaries aim to infer sensitive information from the model's outputs or intercepted data. Their goal is to breach user privacy through analysis and aggregation of data, exploiting patterns to infer protected information. The adversaries, who may have access to intermediate values computed during the diffusion process, could infer information about the input data or the model's parameters.

In our threat model, we assume that the servers hosting the diffusion models operate under an \emph{``honest but curious"} framework. Although these servers faithfully execute the model's computations without tampering with the process or outcomes, they exhibit an intrinsic interest in the users' prompt inputs and generated outputs. This curiosity could lead to privacy breaches if the servers attempt to analyze or infer user data beyond the intended scope of interaction. Specifically, our threat model is based on the following assumptions: (1) The pre-trained diffusion models are accessible to servers in a white-box setting. (2) Prompt texts and output images may contain sensitive information. (3) The embedding model for prompts is selected and could be finetuned by users, which is not available to the servers. We regard these assumptions as realistic under typical operational conditions.

This scenario underscores the need for robust privacy-preserving mechanisms that safeguard user inputs and outputs against both external attackers and potentially intrusive but non-malicious server operators, ensuring that the confidentiality and integrity of the data are guaranteed throughout the diffusion model's processing pipeline.

\section{\system Framework}

In the evolving landscape of generative modeling, particularly within the framework of stable diffusion models, the imperative of establishing privacy-preserving mechanisms is paramount. Prior to the instantiation of such mechanisms, a rigorous examination of sensitive information is warranted. From the point of user interaction, this encompasses primarily the text prompts and the image outputs. Given that the pre-trained model is provided by the server, its associated model weights are not considered sensitive information.

Delving into the mechanics of the stable diffusion model, we encounter an iterative sampling process, which is divided into two stages—forward propagation and subsequent denoising. These stages are presented in Figure~\ref{diffusion}. It becomes evident that the intermediate image, expressed as \( x_i \), converges incrementally towards the final output with each iteration. This convergence bestows upon \( x_i \) the status of sensitive information, necessitating the invocation of privacy protection measures to guard sensitive information.

A meticulous examination of the data flow within the forward and denoising stages reveals that the privacy of the intermediate image \( x_i \) is crucial for the privacy-preserving framework. The text prompts are used by the model's forward propagation phase, which need to be protected. The denoising phase, represented as $x_{i+1} = f(\alpha_i, e_i, x_i)$, employs a triad of inputs—comprising the noise information  $\alpha_i$ , the noise-laden image \( e_i \), and the aforementioned intermediate image \( x_i \)—to engender the next iteration's intermediate image \( x_{i+1} \). Ensuring the privacy of \( x_i \) guarantees the confidentiality of \( x_{i+1} \), thus permitting \( \alpha_i \) and \( e_i \) to be accessible without undermining data privacy.

Moreover, the noisy image, characterized by its stochastic composition, lacks exploitable content. Alone, the noise parameters do not provide enough information to reconstruct \( x_i \). Therefore, protecting the intermediate images and prompts is crucial for ensuring the privacy of the generative process, as they contain the sensitive data across the sampling iterations. In the following sections, we describe how \system protects this information. 

\subsection{Input Protection}

Securing the privacy of prompts and intermediate images \( x_i \) is crucial during the sampling process. However, merely concealing these elements throughout the process is neither time-efficient nor practical.

From a time-cost perspective, the absence of GPU support in existing HE libraries, such as SEAL~\cite{sealcrypto} and TenSEAL~\cite{tenseal2021}, necessitates reliance on CPU processing for encrypted operations. To assess the impact of encryption on processing times, we conducted an experiment encrypting both prompts and intermediate images. Given the substantial computational overhead associated with HE, we extrapolated the time costs for the HE model's forward phase based on the overhead ratios observed in General Matrix Multiplication (GEMM) operations of similar dimensions, which predominantly determine the forward phase's duration. Our findings, as detailed in Table~\ref{tab:estimate}, reveal a significant disparity in processing times between GPU-accelerated and CPU-based computations, as well as between encrypted and plaintext operations. These results underscore the challenge of integrating HE into the model's forward phase without incurring prohibitive time costs, and indicate that innovative solutions are needed to reconcile encryption's security benefits with the practicalities of computational efficiency.

\begin{table}
  \caption{\textbf{Estimated HE-inference time cost. To perform a complete encrypted sampling process of a diffusion model is impractical. $256\times128\times256$ encrypted general matrix multiplication is used as a benchmark to estimate the encrypted inference runtime. Experiments are based on TenSeal~\cite{tenseal2021}, tested on one A100 GPU and one Intel(R) Xeon(R) Gold 6248R CPU (24 cores), respectively. GPU runtime of encrypted model forward is estimated by the result of \cite{jung2021over}. Estimated results are marked by ($\sim$).}}
  \label{tab:estimate}
  \begin{tabular}{ccccc}
    \\
    \toprule
    Case & GPU & CPU & HE-GPU & HE-CPU\\
    \midrule
    $256\times128\times256$ & 0.01s & 0.12s & 67.32s & 932s \\
    Model Forward & 35s & 881s & $\sim$5.72d & $\sim$79.19d\\
    \bottomrule
  \end{tabular}
\end{table}

Incorporating HE within the stable diffusion model, which fundamentally comprises UNet~\cite{ronneberger2015u} and Attention~\cite{vaswani2017attention}, presents a significant challenge due to the prevalence of non-linear operations like softmax, layer normalization, and ReLU. The inherent limitation of HE schemes to only support addition and multiplication operations means that executing these non-linear functions requires them to be approximated by polynomial functions.

Recent approaches involve performing polynomial approximation through Taylor expansions~\cite{vincent2015efficient, ao2023autofhe} or specifically designed linear function~\cite{lu2021soft}. However, this method is prone to error accumulation. The accuracy of Taylor series expansions heavily depends on the polynomial coefficients, which are themselves reliant on data distribution parameters like mean and standard deviation. Given the encrypted nature of the data, dynamically accessing these parameters becomes challenging. Servers might resort to utilizing pre-computed parameters, but this approach is generally effective only in the initial layers of the model due to its static nature. Even with meticulously chosen coefficients that yield satisfactory approximations for individual non-linear operations, the cumulative effect of approximation errors becomes significant as the number of non-linear functions increases throughout the model's depth. 

\begin{figure}[t]
\centerline{\includegraphics[width=0.50\textwidth, height=0.22\textwidth]{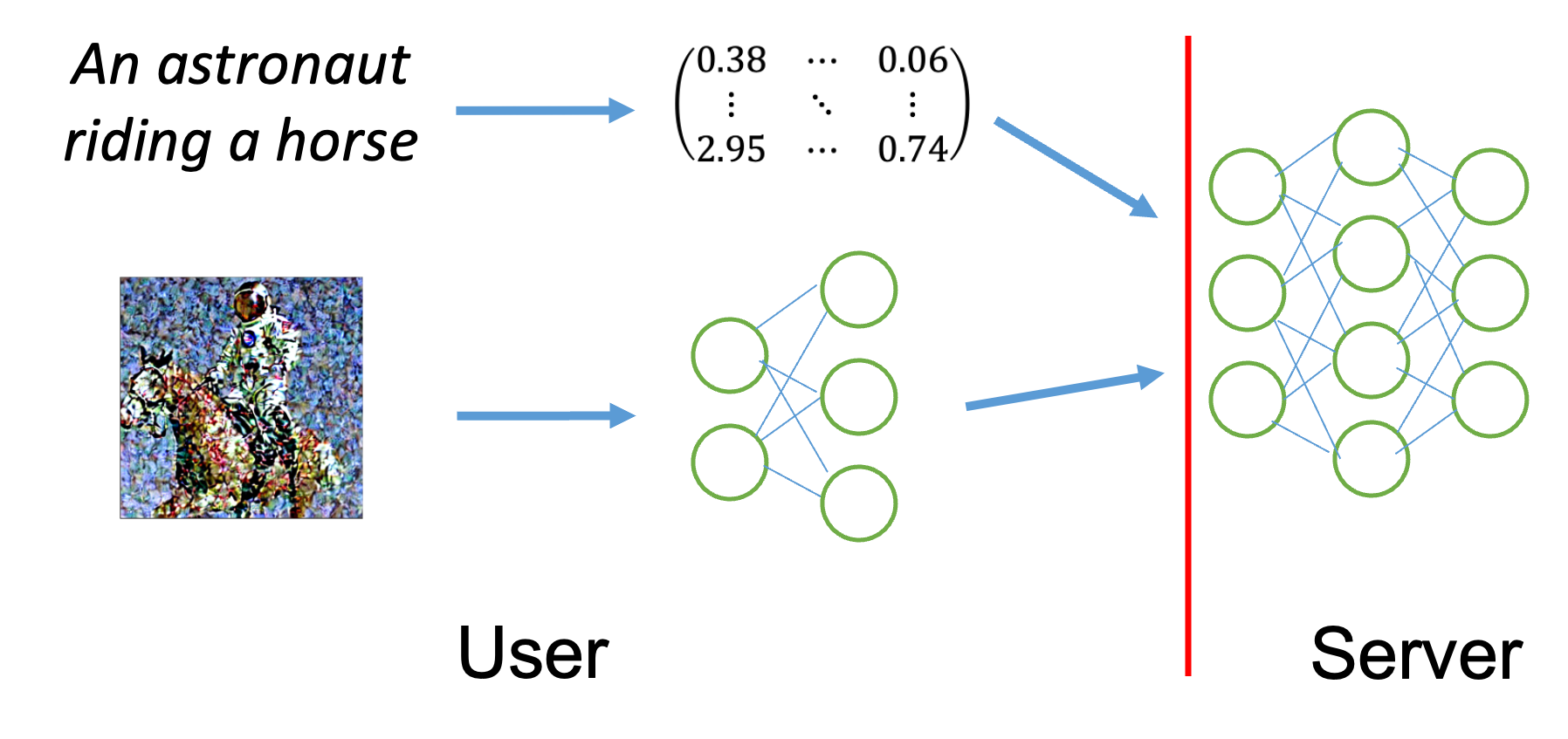}}
\caption{\textbf{Overview of our input protection scheme. Image $x_i$ will not be directly uploaded to the server. The first few layers will be executed by users. Then, the  server performs the rest of computation with the conditioning tensor and the intermediate result.}}
\label{forward}
\end{figure}

To conduct plaintext model forwarding appears to be a feasible solution, while keeping the sensitive information inaccessible by the server. Figure~\ref{forward} shows the model forwarding process.
In detail, the prompts are first encoded into a large conditioning tensor by condition stage model (embedding model), which can be independent of the diffusion model. Users need to choose a proper embedding model as the text encoder. The embedding model can be finetuned for certain tasks. The details of the embedding model are unknown to others including the server. 
By keeping their encoding methods secret, users can prevent servers from reconstructing prompts. Note that existing attacks~\cite{song2020information, feyisetan2021private, chen2024text} typically assume that attackers have access to the embedding model, which is different from the scenario considered in this paper. Considering that the mentioned attacks do not work, the high-dimensional and inherently lossy nature~\cite{tishby2015deep} of this encoding complicates any attempt to directly reconstruct the prompts from the conditioning tensors.

The intermediate image $x_i$ also needs protection. We achieve the protection of $x_i$ by conducting the first few layers on the users' local machine. Ryffel et al.~\cite{ryffel2019partially} have demonstrated that, if the output is not sensitive, we can protect input by keeping the first few layers secret. 
Following this insight, we simply allow users to conduct these calculations locally. To further enhance privacy, users can incorporate different techniques such as adversarial training.

\subsection{Output Protection}
The output image is related to the denoising phase. Operations in this component are element-wise addition and multiplication. Since element-wise operations are not supported by GPU tensor cores, denoising accounts for over 50\% time cost on GPUs. According to the discussion above, we only need to protect $x_i$ and $x_{i+1}$. The basic implementation is shown by Algorithm~\ref{alg1}, where $c_1$ to $c_4$ are scalars, $e_i$ and $x_i$ are tensors with same shape. 

\begin{algorithm}
	\caption{Denoising}
	\label{alg1}
	\renewcommand{\algorithmicrequire}{\textbf{Input:}}
	\renewcommand{\algorithmicensure}{\textbf{Output:}}
	\begin{algorithmic}
		\REQUIRE Noisy Image $e_i$, Noise information $c_1, c_2, c_3, c_4$, Image $x_i$
            \STATE \texttt{/* Current prediction for x0 */}
		\STATE $pred\_x0 = (x_i - c_1 * e_i) / sqrt(c_2)$
		\STATE \texttt{/* Direction pointing to $x_i$ */}
            \STATE $dir\_xi = sqrt(1 - c_3 - c_4^2) * e_i$
            \STATE \texttt{/* Generate noise */}
            \STATE $noise = c_4 * noise\_like(x.shape)$
            \STATE \texttt{/* Previous step of the image */}
            \STATE $x_{i+1} = sqrt(c_3) * pred\_x0 + dir\_xt + noise$
		\ENSURE Image $x_{i+1}$
	\end{algorithmic}
\end{algorithm}

In Algorithm~\ref{alg1}, the image $x_i$ will first be combined with $e_i$ to make a prediction of $x_0$. Accurate $pred\_x0$ serves as a reference point for the model. A direction tensor pointing to $x_i$ is calculated to show how model should adjust the current noisy $x_i$. Then, we get $x_{i+1}$ from the combination of $pred\_x0$, $dir\_x$ and a random noise.

\begin{figure*}[htbp]
\centerline{\includegraphics[width=0.90\textwidth, height=0.28\textwidth]{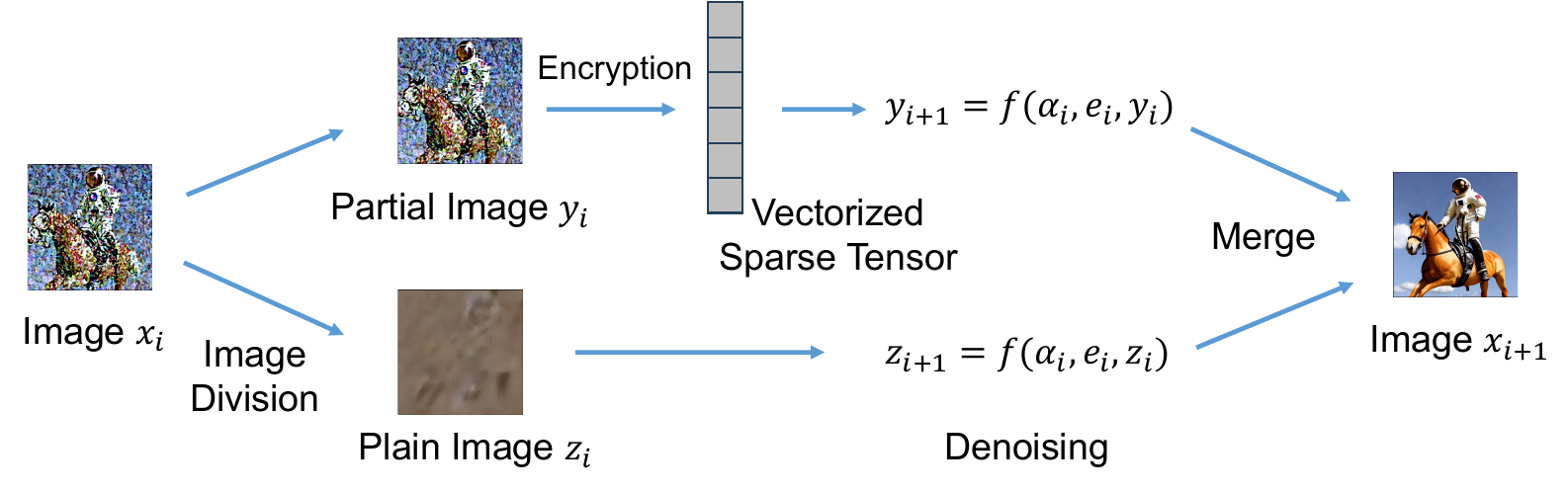}}
\caption{\textbf{Overview of our partial encryption scheme. $x_i$ is divided into $y_i$ and $z_i$, with most of information kept by $y_i$. Then $y_i$ is encrypted while $z_i$ remains unencrypted. After denoising, $x_{i+1}$ can be constructed by merging $y_{i+1}$ and $z_{i+1}$.}}
\label{partial}
\end{figure*}

Initiating encryption at the start and concluding with decryption might seem straightforward, yet it incurs inefficiencies due to the necessity for re-encryption and redundant ciphertext calculation. Particularly, since both \( pred\_x0 \) and \( x\_prev \) are in ciphertext form and their associated computations are complex, certain operations might need to be executed in the ciphertext domain due to the sequence of computations. To streamline this process, the aim is to maximize the number of operations conducted in plaintext, resorting to ciphertext computations only when absolutely necessary. Recalling Algorithm~\ref{alg1}, we need to merge all the plaintext computations and conduct operations corresponding to $x_i$ as late as possible. Through strategic reorganization of the computational steps, we can significantly diminish the need for ciphertext operations to merely a single tensor multiplication and addition. Moreover, the advantage of element-wise operations should be utilized, i.e., vectorization. Without the need to consider about the shape of tensors, the sparse values can be encrypted by lightweight vectors in the HE libraries. Though it reduces the multiplication depth, it brings great performance improvement. 

Given that \( x_i \) remains constant throughout the model's forward pass, the denoised image from the preceding step can be directly utilized as the input for the subsequent iteration. To address the constraints imposed by the multiplication depth, relinearization techniques should be applied intermittently. However, due to limited support for relinearization in existing libraries~\cite{tenseal2021}, our implementation temporarily adopts infrequent re-encryption as an alternative strategy.

The refined encrypted denoising algorithm is illustrated in Algorithm \ref{alg2}. Some parameters such as $pred\_x0$ are not explicitly calculated. The plaintext computations such as calculating $dir\_xt$ and noise generation are unmodified, while the computation involved by $x_i$ is compressed to include only one ``multiply" and ``add" operation. Here, \( x\_prev \), the encrypted version of the intermediate image, is leveraged in the denoising process for the next iteration, denoted as \( enc_{x_{i+1}} \). This encrypted intermediate image is also transmitted back to the user for decryption, enabling them to compute the initial layers of the model's forwarding phase. As the intermediate images remain encrypted on the server throughout the process, the privacy of the output is effectively maintained. 

\begin{algorithm}
	\caption{Encrypted Denoising}
	\label{alg2}
	\renewcommand{\algorithmicrequire}{\textbf{Input:}}
	\renewcommand{\algorithmicensure}{\textbf{Output:}}
	\begin{algorithmic}
		\REQUIRE Noisy Image $e_i$, Noise information $c_1, c_2, c_3, c_4$, Encrypted Image $enc_{x_i}$
		\STATE \texttt{/* Direction pointing to x\_i */}
            \STATE $dir\_xi = sqrt(1 - c_3 - c_4^2) * e_i$
            \STATE \texttt{/* Generate noise */}
            \STATE $noise = c_4 * noise\_like(x.shape)$
            \STATE \texttt{/* Multiplication factor */}
            \STATE $factor = sqrt(c_3 / c_2)$
            \STATE \texttt{/* Addition part */}
            \STATE $add\_part = dir\_xt + noise - factor * c_1 * e_i$
            \STATE \texttt{/* Previous step of the image */}
            \STATE $x\_prev = factor * enc_{x_i} + add\_part$
            \STATE \texttt{/* sent x\_prev to user and decrypt */}
            \STATE $x_{i+1} = x\_prev.decrypt()$
		\ENSURE Image $x_{i+1}$, Encrypted Image $x\_prev$
	\end{algorithmic}
\end{algorithm}


\section{Partial Encryption by Distortion} \label{partial}
Manipulating image is expensive even in a latent space. According to Algorithm~\ref{alg2}, in each iteration, we conduct element-wise multiplication and addition for a tensor, which means thousands of ciphertext-plaintext multiplication should be performed. To mitigate these computational demands, we introduce a strategy combining partial encryption with ciphertext-plaintext hybrid computation.

Partial encryption for images~\cite{van2004partial} has long been early regarded as a powerful tool to reduce the encryption cost. In the context of HE, there are very few studies~\cite{yonetani2017privacy} leveraging partial encryption. The reasons may be manifolds: (1) Partial encryption engenders complex hybrid data structures and computations, which are not readily accommodated by existing HE libraries. (2) The sparsity induced by partial encryption contradicts the prevailing trend in hardware development, which favors dense computational paradigms exemplified by tensor cores and SIMD (Single Instruction, Multiple Data) architectures. (3) Employing partial encryption necessitates a delicate balance between security and computational efficiency, as it poses potential risks for information exposure. Despite these challenges, we advocate for the integration of partial encryption with HE, arguing that the benefits of hybrid computation—specifically in reducing computational and memory burdens—outweigh the concerns. 
In the subsequent sections, we will demonstrate that partial encryption serves as an ideal complement to HE. By strategically encrypting only certain components of the data, we can significantly reduce computational overhead without sacrificing security, thereby enhancing the efficiency and practicality of HE in real-world applications. This approach leverages the strengths of both partial encryption and HE, offering a balanced solution that maintains data privacy and security while optimizing performance. 

\subsection{Partial Encryption Scheme}

Element-wise operation suits partial encryption well, since it can gain linear time saving with the reduction of encrypted elements, and it is convenient for data decomposition. We conceptualize the image \( X \) as a composite of two components: \( X = Y + Z \), where \( Y \) encapsulates the bulk of \( X \)'s information, and \( Z \) contains elements deemed less critical. This decomposition ensures that \( Y \) and \( Z \) are mutually exclusive, with \( Y \times Z = 0 \), indicating that each element of \( X \) is exclusively present in either \( Y \) or \( Z \), with the counterpart being zero. In this framework, HE is applied to \( Y \), while \( Z \) remains in plaintext, which is written as follows. 
\begin{equation}
    f(X) = f(Y) + f(Z).
\end{equation}

Figure~\ref{partial} describes this scheme in detail. It is unnecessary to amalgamate \( y_{i+1} \) and \( z_{i+1} \) at every iteration; they can be computed independently, given the constraints on multiplication depth. The main challenge lies in minimizing the encrypted portion \( Y \) to the smallest possible subset while retaining the majority of the image's information. Notably, this approach to partial encryption preserves the integrity of the computational results without resorting to approximation, allowing for adjustable levels of security through different decomposition techniques and parameter configurations.

After removing data points, elements of both of $Y$ and $Z$ are discretely distributed. Thus, sparse computation can be applied. Due to its superiority in reducing time complexity and minimizing memory requirements, sparse computation has consistently been at the forefront of research and development. This approach, which emphasizes the processing of data structures where the majority of elements are zeros, enables more efficient storage and faster computational operations. In the context of plaintext-ciphertext mixed operations, sparsity has more chance due to the extreme time cost gap between encrypted and unencrypted operations. High overhead of HE operations forces us centralize computing resources to handle a few operations, which alleviates efficiency problem of discrete computation. 

\subsection{Image Division}
 
In this work, we use additive distortion to remove unimportant data points from image tensors. The central advantage of distortion is that it works in the latent space, where most computation happens. Latent computation limits most of pixel-based methods like semantic segmentation.

According to the definition of additive distortion, if we remove $n$ points from image $X$ with indices $(i_1, j_1), (i_2, j_2), \dots, (i_n, j_n) \in I$ and generate a new image $Y$, the additive distortion can be written as:
\begin{equation}
    D(X, Y) = \sum_{(i,j) \in I} \rho_{ij} |x_{ij}|.
\end{equation}
This equation implies that the order of removing two points will not affect the calculation. As $\rho$ is only related to $X$, it can be pre-calculated and treated as a constant matrix. Then, every data point in $Y$ contributes independently to the whole additive distortion. Calculating distortion after removing a point from $Y$ is simply adding a term corresponding to the data point. Thus, distortion of every point can be discussed separately.

Moreover, additive distortion can be considered as some kind of information. Distortion of a point can be treated a quantified form of information it carries. A direct way is to iteratively remove points with low distortion. However, the iteration should stop before important information leaks. Therefore, it requires a threshold for terminating the iteration. We can calculate the total distortion while all points are removed as follows:

\begin{equation}
    Whole\_D(X) = \sum_{i,j} \rho_{ij} |x_{ij}| = D(X, 0),
\end{equation}

With a total distortion, the importance of a point can be quantified by the ratio between its own distortion and the total distortion. This provides convenience for us to set a threshold for points removing for different images. Algorithm~\ref{alg3} shows an overview of this procedure. First, the cost matrix is generated by cost function and $X$. Then, it conducts element-wise multiplication with $X$ to create the distortion matrix. The total distortion can be calculated incidentally. We iteratively find the minimum value in the distortion matrix and set the element at the corresponding position of $Y$ to 0. Since this brings distortion, we then update distortion and the matrices. When the distortion caused by removed points reaches the threshold, the algorithm stops and output the sparse image $Y$. In this algorithm, cost\_function can be chosen according to the needs. The remaining tensor $Z$ can be simply generated by $Z = X - Y$ after the removal.

\begin{algorithm}
	\caption{Point Removal (Basic)}
	\label{alg3}
	\renewcommand{\algorithmicrequire}{\textbf{Input:}}
	\renewcommand{\algorithmicensure}{\textbf{Output:}}
	\begin{algorithmic}
		\REQUIRE Image $X$, $threshold=0.01$, $distortion=0$, $Y=X$
            \STATE \texttt{/* Get cost matrix*/}
		\STATE $C = cost\_function(X)$
		\STATE \texttt{/* Distortion matrix */}
            \STATE $D = C * X$, $Whole\_D = D.sum()$
            \STATE \texttt{/* Iteratively remove points */}
            \REPEAT 
            \STATE \texttt{/* Get min-distortion element */}
            \STATE $min\_ele = argmin(D), Y[min\_ele] = 0$
            \STATE \texttt{/* Update distortion */}
            \STATE $distortion \mathrel{+}= D[min\_ele], D[min\_ele] = inf$
            \STATE \texttt{/* End loop when reach threshold */}
            \UNTIL{$distortion / Whole\_D < threshold$}
		\ENSURE Sparse Image $Y$
	\end{algorithmic}
\end{algorithm}

This implementation is quite inefficient, due to the existence of a large number of iterations. Considering that this procedure should be frequently executed during stable diffusion sampling, we propose the following methods for further speedup.

We first check some conditions and conduct batch deletion. The maximum threshold in the procedure is $dis\_threshold = Whole\_D * threshold$. Then in the process, remaining distortion can be defined as $dis\_remain = dis\_threshold - distortion$. All points meets this inequality can be removed together:
\begin{equation}
    distortion * remain\_points < dis\_remain,
\end{equation}
where remain\_points denotes to the number of points not removed. This condition can be simply done by function `where' in torch or numpy. 

Other functions such as `topk' also helps for acceleration. While these optimizations are utilized, time cost of the procedure can be reduced from several hours to less than 5s.

\begin{figure*}[htbp]
\centerline{\includegraphics[width=0.90\textwidth, height=0.32\textwidth]{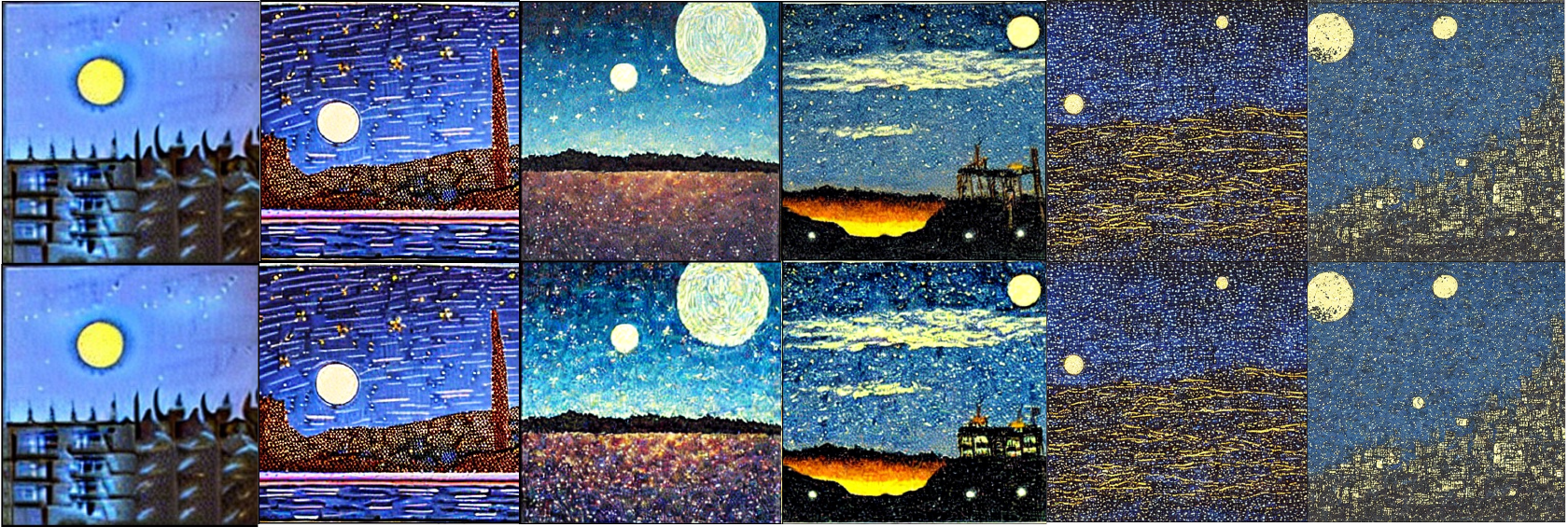}}
\caption{\textbf{Examples to compare the generated images from the plaintext version and the encrypted version. Top row: samples from stable diffusion model. Bottom row: samples from \system model (Prompts: "Night Scene with 2 moons." The height and width of the images are equally and sequentially set as \{256, 384, 512, 768, 1024, 1280\}. The images corresponding to these size are listed from left to right. All other parameters are equally set).}}
\label{enc_acc}
\end{figure*}

\subsection{Sparse Encryption and Computation}
Considering that tensor $Y$ contains only partial data point of $X$, and these points are discontinuous distributed in the tensor, its sparsity can be utilized for acceleration. Sparsity has two primary advantages: (1) Tensor $Y$ has thousands of elements (zeros and nonzeros), their ciphertexts account for large space. (2) Conducting encryption and calculation only on nonzero elements save a substantial amount of time, and embodies the meaning of partial encryption. $Z$ can keep dense since $f(Z)$ is efficient enough so that the payoff from optimization is small.

Special workflow leads to customized operations and data structure. $f(Z)$ is calculated in dense format. Considering that there are only element-wise operations in $f$, and elements of $X$ are separated in $Y$ and $Z$, respectively, $f(Z)+ f(Y)$ can be done without calculation. We can simply perform assignments as $Z[Y.indices] = Y.values$ to merge these two tensors together. Addition and multiplication should also be specifically designed. The $factor$ and $add\_part$ in Algorithm~\ref{alg2} are all dense tensors. Since $f(Z)$ deals with partial computation, $f(Y)$ should only focus on the elements in $Y.indices$. That means we do not need to explicitly transform $factor$ and $add\_part$ to sparse tensor. Partial addition can be performed as:
$$Y.values \mathrel{*}= factor[Y.indices],$$
$$Y.values \mathrel{+}= add\_part[Y.indices].$$
Traditional sparse tensor or sparse matrix format does not directly support these functions. Previous work has used sparse matrix multiplication or tensor manipulation under HE~\cite{chen2021homomorphic, yang2022fpga}. However, sparse HE still lacks support for operations that facilitate hybrid computation between sparse and dense matrices.

Here, we leverage coordination (COO) format to store the sparse tensor. Compared with compressed format like compressed sparse row (CSR), compressed sparse column (CSC), COO needs more memory for indices. But the advantages of COO is its flexibility and simplicity. Moreover, its indices can be directly accessed by dense tensor without any process, which provides convenience for sparse-dense hybrid computation. Since memory cost of encrypted values is much larger than plaintext indices, the memory advantage of compressed formats is almost negligible. 

Same as plaintext sparse tensor, only nonzero elements are stored in the encrypted COO sparse tensor. This helps to reduce much time in the context of the rapid increase in encryption time with the number of elements. We provide interface to perform encryption on COO sparse tensor with different encryption schemes, such as BFV~\cite{brakerski2012fully} and CKKS~\cite{Cheon2017HomomorphicEF}. The encrypted values could also be organized as a vector for vectorized operation, in order to gain higher performance by SIMD encoding\cite{smart2014fully}. Since performing encryption and computation becomes much cheaper, and the number of operations is largely reduced by Algorithm~\ref{alg2}, we will run reencryption when the accuracy is on the decline.

\section{Evaluation}
In this section, we will conduct a comprehensive evaluation of \system, examining its performance across three critical dimensions: accuracy, performance, and security. In the accuracy evaluation, we mainly focus on whether HE brings precision loss. For performance evaluation, we will analyze the system's runtime under various parameter configurations, providing insights into how different settings influence overall efficiency. This analysis will extend to examining the system's responsiveness in scenarios with varying levels of data sparsity. The security dimension of our evaluation focuses on the challenges and vulnerabilities introduced by partial encryption. We will delve into the potential for information leakage, and offer a detailed discussion on how partial encryption, while beneficial for reducing computational overhead, could expose the system to certain security risks. 

\subsection{Implementation Details and Settings}
The experimental results are generated by stable diffusion model v1.4 with a PLMS sampler to generate images at a default size. Our implementation is based on the open-source implementation of stable diffusion models~\cite{latentdiffusion}.

The HE component in our work uses CKKS~\cite{Cheon2017HomomorphicEF} scheme by TenSEAL~\cite{tenseal2021}. We choose the poly modules degree of 8,192 and the coeff-modules of 26. For a simulation purpose, users and server were both implemented in a single computer and thus no transmissions among them were considered in the experiments. All parameters are kept same in the plaintext version and the encrypted version.

In this work, the computation resource are a NVIDIA A100-PCIE GPU with 40GB memory and a Intel(R) Xeon(R) Gold 6248R CPU running at 3GHz.

\subsection{Accuracy}

Though our methods do not explicitly include polynomial approximation, there will be loss of accuracy due to the approximate encryption scheme~\cite{Cheon2017HomomorphicEF}. Here, we utilize five different methods to evaluate the quality of the generated images. Cosine similarity and Min squared error (MSE) are basic mathematical tools to directly compare the data distribution. In the pixel space, Structure Similarity Index Measure (SSIM), Peak Signal to Noise Ratio (PSNR)~\cite{hore2010image} work in different aspect. SSIM integrated luminance, contrast and structure to fit the intuitive feeling of the human eye. PSNR stands at the signal side to evaluate the power of a signal and the power of corrupting noise. KL divergent comes from cross entropy, which quantifies the redundant information. We refer to these five indicators, in order to make an overall judgment.
\begin{table}
  \caption{\textbf{Different similarity measures between the images generated by plaintext version and encrypted version. All parameters and prompts are equally set. The results in the table are the mean value of the generated samples. }}
  \label{tab:similarity}
  \begin{tabular}{ccccc}
  \\
    \toprule
    Cosine & SSIM & PSNR & MSE & KL Divergence\\
    \midrule
    0.9818 & 0.9995 & 68.65 & 0.0089 & 0.0046\\
    \bottomrule
  \end{tabular}
\end{table}

As Table~\ref{tab:similarity} shows, our homomorphically encrypted diffusion inference can generate very similar images compared with the original unencrypted one. According to the table, we have PSNR > 40dB, which means the signal is absolutely dominant compared with the noise in the encrypted data. High cosine similarity and low KL divergence shows that the two figures are very close to each other. Furthermore, cosine similarity is lower than SSIM. That implies the encrypted images show some differences with the original ones, but in the context of structure and perceived quality, the encrypted images keep most of information.  Figure~\ref{enc_acc} provides a more intuitive example for different image sizes. We utilize identical random seeds to generate images in both the encrypted and unencrypted versions, maintaining the exact same settings. Although there are minor pixel variations, the generated images are nearly identical. The MSE value is exceptionally low, indicating that the two images are element-wise similar, rather than just broadly similar.


\subsection{Performance}

In the performance evaluation, our primary focus is on determining the feasibility of our HE-Diffusion model and understanding the impact of sparsity on its functionality. Table~\ref{tab:time} provides the process time of different algorithms. To the best of our knowledge, a homomorphically encrypted stable diffusion model does not exist. Therefore, our comparisons are made between the plaintext version and our various algorithms. Due to the inevitable introduction of overhead of security, the plaintext version is the fastest. And just as we expected, simply conducting unoptimized homomorphic encryption brings significant complexity. However, after optimization and our sparse design, this overhead is largely reduced. Based on our findings, our most efficient sparse implementation achieves a computational speed that is on par with the plaintext version in terms of order of magnitude, with 500 times speedup compared with the naive encrypted implementation. This significant achievement underscores the practicality of our approach for users, as \emph{it ensures that the enhanced privacy provided by homomorphic encryption does not come at the expense of performance.}

In Table~\ref{tab:time}, the time costs of GPU acceleration and pure-CPU follow different mode of change. With customized design for matrix multiplication, GPU can reduce the model forward time from 8.8s to less than 0.04s per iteration. As a result, the distribution of computation hotspot is quite different. In pure-CPU version, model forward accounts for the primary time costs even after homomorphically encrypted denosing. However, in the GPU version, denosing is the most computational intensive component. We also considered the time cost of communication. Since our protect scheme needs users and server to exchange data, we evaluated the time of all communications. According to Table~\ref{tab:comm}, uploading conditional tensors takes negligible time, since it is only performed once. Weights and encrypted images should be transmitted every iteration, and account for less than $0.024s$. Compared with the encrypted denoising process, it only brings 3\% extra time cost. As a result, we can confirm that the additional communication overhead does not significantly impact the runtime performance.

\begin{table}
  \caption{\textbf{Exact time cost of one denoise process and the whole \system inference. In the `GPU' version, GPU is applied to accelerate model forward, while `CPU' denotes to a pure CPU implementation. `Plain' refers to Algorithm~\ref{alg1}. `Enc' simply encrypt $x\_i$ for Algorithm~\ref{alg1} with nothing else changed. `Enc\_opt' refers to Algorithm~\ref{alg2} with less frequent reencryption. `Sparse' is the sparse version described in Section~\ref{partial} (i.e., the final version of \system). Here, reencryption is applied instead of relinearization. All parameters of stable diffusion are set as default values. In the sparsity test, we choose threshold=$0.01$ as a conservative parameter setting.}}
  \label{tab:time}
  \begin{tabular}{ccccc}
    \\
    \toprule
    Case & Plain & Enc & Enc\_opt & Sparse\\
    \midrule
    Denoise once & \textbf{0.12s} & 354.34s & 1.67s & \textbf{0.70s} \\
    Inference (CPU) & \textbf{0.24h} & 10.06h & 0.41h & \textbf{0.32h}\\
    Inference (GPU) & \textbf{35s} & 9.84h & 203s & \textbf{106s} \\
    \bottomrule
  \end{tabular}
\end{table}

\begin{table}
  \caption{\textbf{Estimated communication time cost of the \system. There are in total three kinds of communication in the entire process. The conditional tensor (default size: $3\times 77\times 768$) will only be uploaded once before iterations. Intermediate weights and encrypted images have the same shape of the original images (default size: $3\times 4 \times 64\times 64$), while ciphertexts will be SIMD-encoded. Time cost of uploading and downloading encrypted images are counted together. We assume that users transmits data with Internet bandwidth below 10MB/s.}}
  \label{tab:comm}
  \begin{tabular}{cccc}
    \\
    \toprule
    Case & Conditional Tensor & Weights & Encrypted Images\\
    \midrule
    Time Cost & 0.017s & 0.0047s & 0.019s\\
    \bottomrule
  \end{tabular}
\end{table}

\begin{figure}[t]
\centerline{\includegraphics[width=0.45\textwidth]{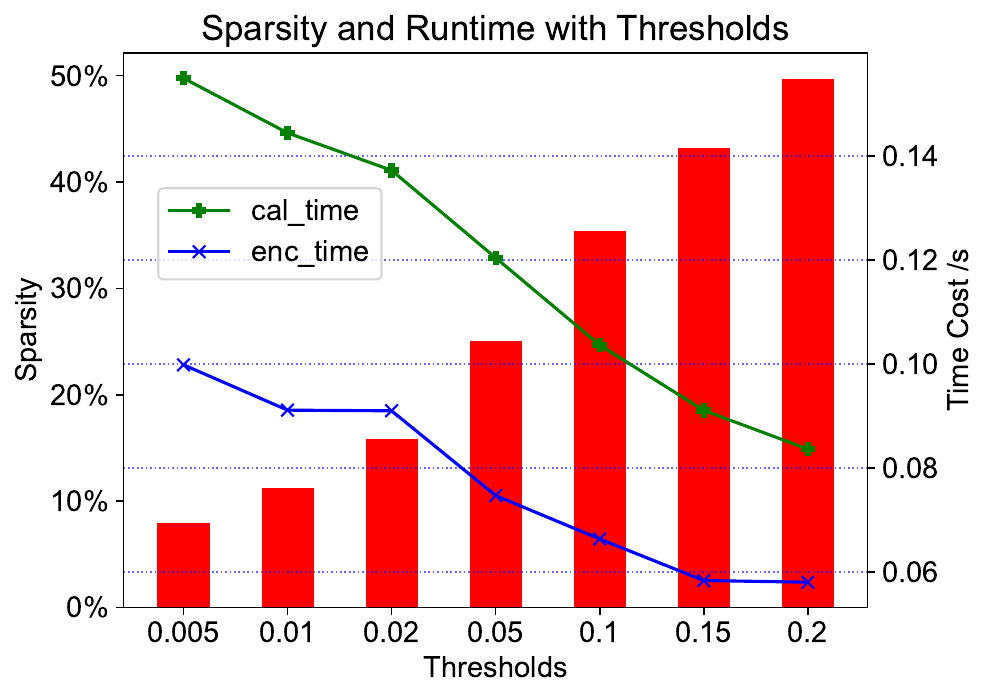}}
\caption{\textbf{Variation of sparsity and the ability to keep sensitive information under different thresholds. Images are tested in the latent space. The red bars denotes to the percentage represented sparsity. The blue line and the green line are the cosine similarity between the remained image (encrypted part), the removed part and the original image, respectively.}}
\label{threshold}
\end{figure}

In details, different parameter settings affect the performance results. As shown in Figure~\ref{threshold}, when we set a higher threshold for Algorithm~\ref{alg3}, higher sparsity will be achieved, and the time cost (both the encryption time and the calculation time) will decrease. This is understandable, since high sparsity leads to fewer non-zero elements, which benefits encryption and calculation. However, this is not without cost. The risk and trade-off will be discussed in the next section. 

\begin{figure}[t]
\centerline{\includegraphics[width=0.45\textwidth]{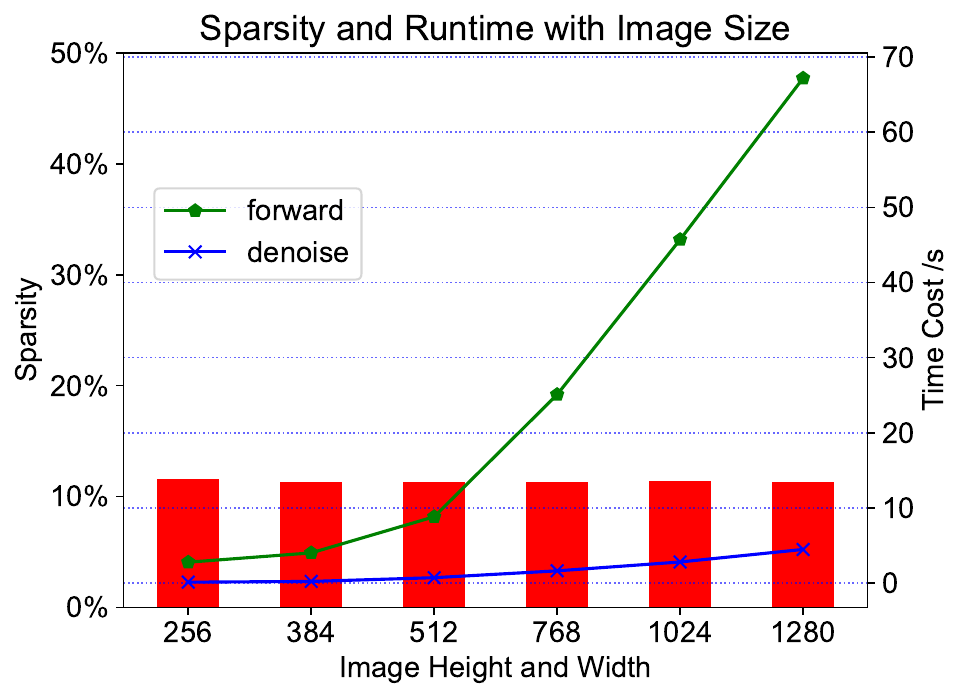}}
\caption{\textbf{Sparsity and time cost of model forwarding and denoising components with the variation of image size. Without loss of generality, height and width are set equally. Threshold is set as 0.01. Experiments are done on CPU due to the memory limitation of GPU.}}
\label{shape}
\end{figure}

The influence of the image size is also considered, as Figure~\ref{shape} shows. With the increase of image size, the time cost of model forwarding and denoising follows Quadratic growth, while sparsity almost keeps unchanged. Since model forwarding accounts for the majority of time, the ratio between it and denoising are growing rapidly when enlarging image size, which means, comparatively speaking,  the total run time will be closer to plaintext (unencrypted) version.

\begin{figure}[t]
\centerline{\includegraphics[width=0.50\textwidth]{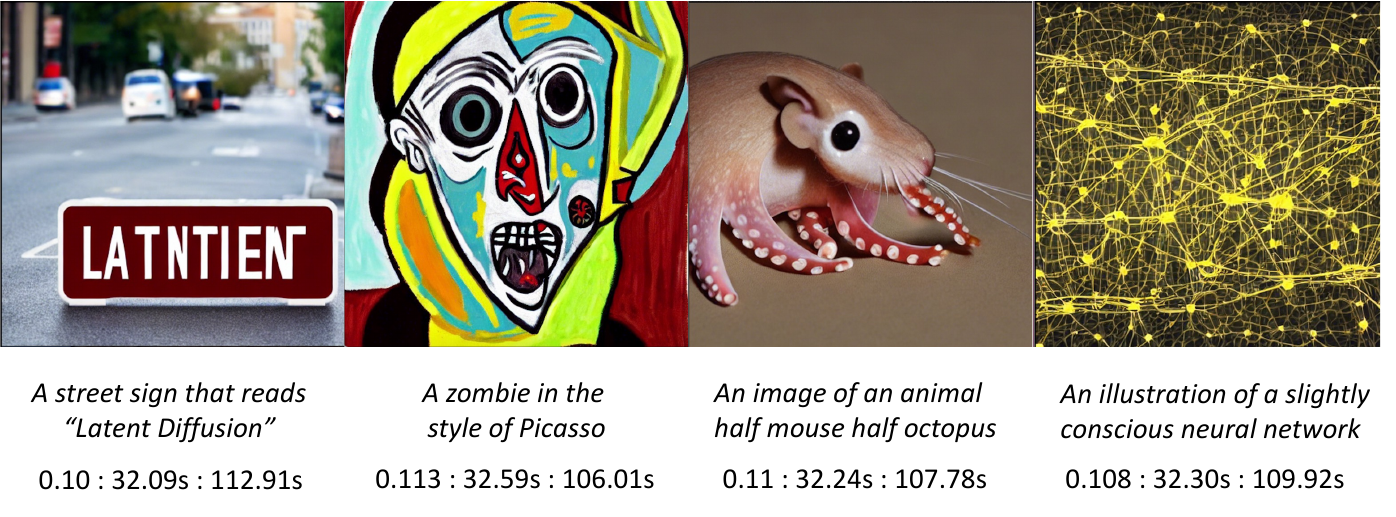}}
\caption{\textbf{Performance results with different prompts.  The time cost results are organized as   ``sparsity : plaintext running time : \system running time". Experiments are conducted on one A100 GPU.}}
\label{prompts}
\end{figure}

Different prompts also influence the performance of \system, as shown Figure~\ref{prompts}. Considering that different prompts are embedded into conditioning tensors with the same shape, there should not be  a significant difference in their time costs. However, conditioning tensors affect the denoising process through noise information. As the information distribution in the intermediate images vary, different sparsity is achieved, which will lead to different runtime performance.

\subsection{Security of Partial Encryption}
Partial encryption is commonly associated with the risk of information leakage. In this section, we will show the information preserving ability of our  point removal algorithm. Since the computation happens in a latent space, our partial encryption also works in this latent space. Similarity metric is utilized to quantify the information leakage. In the latent space, we do not use pixel-based SSIM and PSNR, but we mainly consider cosine similarity and KL divergence. The former can directly compare the two tensors, while the latter helps to reduce bias with a holistic view.
\begin{figure}[t]
\centerline{\includegraphics[width=0.45\textwidth]{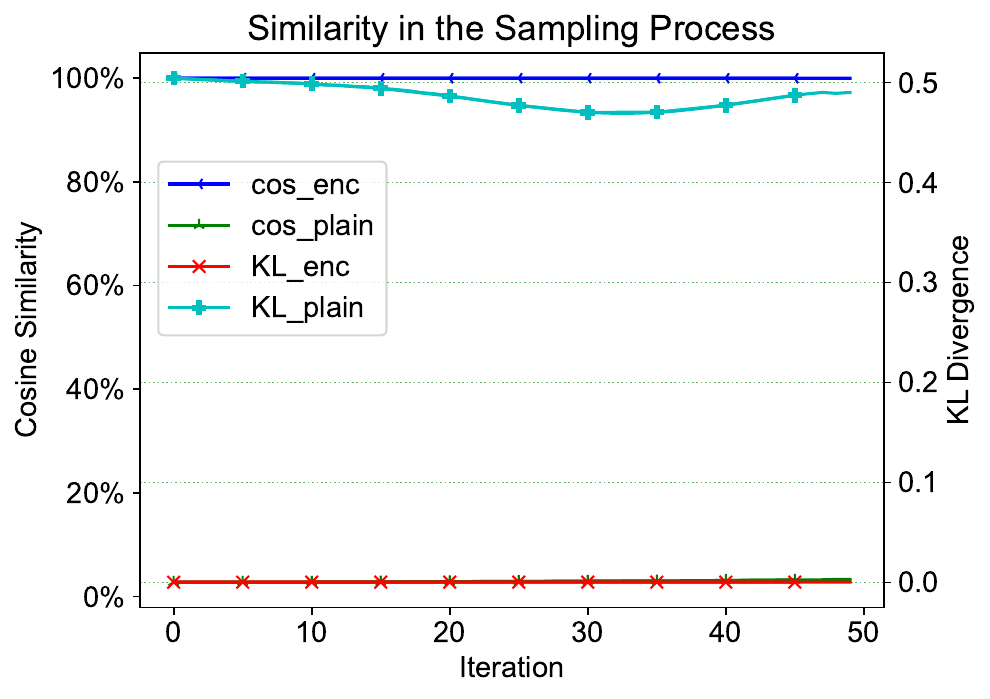}}
\caption{\textbf{Information leakage of partial encryption through the whole sampling process. Cosine similarity and KL divergence between two latent-space images are used to evaluate the effect of information preservation. The blue line and the green line denote cosine similarity. The cosine similarity between the encrypted portion and the original image is approximately 99.96\%, compared to 2.81\% for the plaintext portion. The KL divergences for the encrypted portion are less than 0.0003, whereas they are at least 0.4693 for the plaintext portion. Threshold is set as 0.01.}}
\label{similarity}
\end{figure}

After partial encryption, the original image tensor $X$ is divided into encrypted $Y$ and plaintext $Z$. We evaluate the cosine similarity and the KL divergence $KL(X||Y), KL(X||Z)$ through the entire sampling process. Our experiments primarily focus on determining whether there is any information leakage throughout the entire sampling process and how this potential leakage varies with different parameter configurations.

Figure~\ref{similarity} shows the results in all iterations in a sampling process. Cosine similarity between $X$ and $Y$ is near $100\%$, while $X$ and $Z$ do not act alike. KL divergence provides further confirmation. When $KL(X||Y)$ approaches 0, it means $Y$ represents $X$ very well. 

According to these results, the concern of information leakage can be addressed in two ways: (1) $Y$ keeps most of information from $X$. (2) $Z$ inherits only very small amount of information from $X$, which is far from reconstructing $X$. To make this comparison more intuitive, we transform $X, Y, Z$ from the latent space to the pixel space to gauge the difference.

\begin{figure}[t]
\centerline{\includegraphics[width=0.50\textwidth, height=0.45\textwidth]{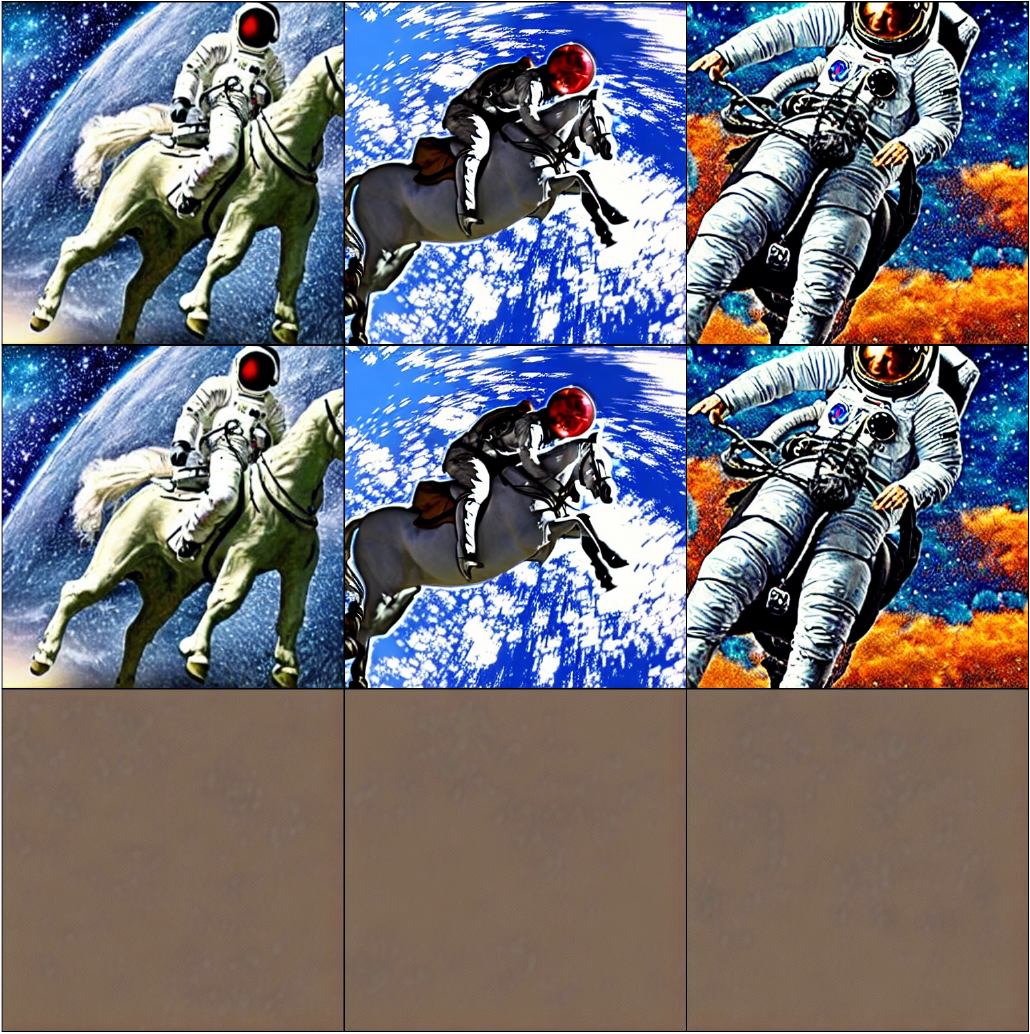}}
\caption{\textbf{Intuitive test for the security of partial encryption.  Prompts: "a photograph of an astronaut riding a horse". Threshold is set as 0.05. The three images above are the original images. The center three are the partial encrypted images. The three images below are the plaintext images. The plaintext images are purely noise.}}
\label{figurecompare}
\vspace{-15pt}
\end{figure}

Figure~\ref{figurecompare} provides an example. We can observe that the partial encrypted figure is very similar with the original figure, while the plaintext figure does not reflect any helpful information. This implies an advantage of performing partial encryption in the latent space that, even if there is some information kept by $Z$, attackers can hardly detect it using structure or other methods in pixel space after transformation.

\begin{figure}[t]
\centerline{\includegraphics[width=0.45\textwidth]{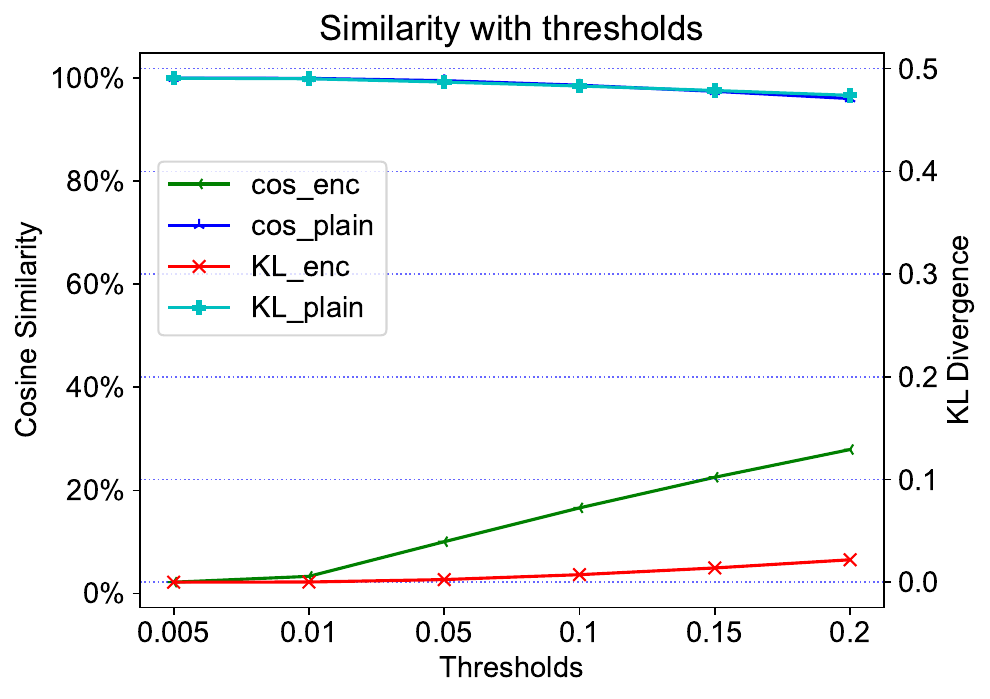}}
\caption{\textbf{Information leakage with different thresholds. Cosine similarity and KL divergence between two latent-space images are used to test the effect of information preserving. With the increase of threshold, the similarity between the encrypted image and the original image will decrease, while the plain image becomes more similar with the original one. Considering that the behavior of similarity keeps stable in each iteration, the last iteration is chosen for the experiments.}}
\label{threshold_cos}
\vspace{-15pt}
\end{figure}

Figure~\ref{threshold_cos} further presents the similarity with different parameter settings. When we set a larger threshold for Algorithm~\ref{alg3}, though higher sparsity and lower time cost will be achieved, there will be higher security risk. According to Figure~\ref{threshold_cos}, with the threshold increasing, cosine similarity between $X$ and $Y$ will decay, and cosine similarity between $X$ and $Z$ will augment. KL divergence shows the same trend. We have to set a proper threshold to achieve sparsity and prevent information leakage simultaneously, which indicates a trade-off between security and efficiency. Considering that the cosine similarity between $Z$ and $X$ begin to increase rapidly from threshold = 0.01, we use 0.01 as the threshold in other experiments for security guarantee.

\section{Discussion}
In this work, we propose \system, a privacy-preserving stable diffusion model. It works by safeguarding the input prompts and output data for users when the computation is hosted by unreliable servers. \system can be utilized across diverse fields where image generation is beneficial, particularly in contexts where the protection of sensitive data is a primary concern. In this section, we present discussions and limitations of \system.

\subsection{Applicable Scenarios}

A fundamental scenario involves users keen on preventing their activities from being monitored and analyzed. The prompts they use and the images generated could reveal personal information and lifestyle patterns, posing a risk to their privacy. In domains such as healthcare, this concern is magnified as inputs often originate from confidential databases, where patient data and sensitive medical information must be handled with utmost discretion. Encryption in such contexts not only safeguards against unauthorized access but also ensures compliance with data protection regulations, thereby maintaining the integrity of sensitive data.

For content creators, the stakes are similarly high. Encrypted outputs can serve as a protective measure against copyright infringement claims, offering a layer of security that shields creators from potential legal disputes. This is particularly relevant in industries where intellectual property rights are rigorously enforced and the line between inspiration and infringement can be thin. By utilizing encryption, creators can maintain control over their work, ensuring that their creations are shared and used in ways that respect their original intentions and copyright.

Moreover, in the realm of digital content creation, encrypted outputs can also prevent unauthorized reproduction and distribution, preserving the value of the content and the creators' revenue streams. In scenarios where content is personalized or contains elements tailored to specific individuals or groups, encryption can prevent the misuse of such content, and ensure that sensitive information is not exploited for malicious purposes.

\subsection{Limitations and Practical Considerations}
Our approach does come with certain limitations. First, we have not implemented full encryption for the forward pass of the model. This necessitates users to meticulously manage the embedding of their prompts and coordinate data exchange with the server, introducing a layer of complexity to the user experience. Additionally, our implementation strategy for denoising leverages specific characteristics unique to this component, which unfortunately limits the applicability of our methods to other models. This specialization means that the insights and efficiencies gained in our denoising process are not readily transferable to different model architectures or applications. Expanding our methodology to accommodate a broader range of models and computational tasks would require reevaluation of our encryption strategies and possibly the development of new techniques to ensure both security and performance across diverse computational contexts.

To make this work more practical in the real world, there are some improvements to be done. 
For ease of implementation, we introduced certain simplifications, which may adversely affect the quality of the generated images and their sparsity.  We used the original version of stable diffusion instead of diffuser~\cite{von-platen-etal-2022-diffusers}. To achieve better images and enable more pre-trained models, diffuser should be supported.  Furthermore, our primary objective was to demonstrate the effective synergy between HE and partial encryption. Therefore, we did not delve into fine-grained optimizations for sparse tensors and sparse operations. Our work is implemented in Python based on TenSeal~\cite{tenseal2021}. More techniques of high performance computing and parallel computing can be applied if we transfer some components to C++. Moreover, the methods to achieve sparsity can indeed vary; in our case, we opted for a straightforward approach by employing HILL~\cite{li2014new}, a simple cost function introduced by Li et al. Adopting a more sophisticated cost function or a more efficient method for collecting sensitive information could significantly enhance sparsity, directly benefiting the overall effectiveness of our approach.


\section{Conclusion}
The integration of HE within the stable diffusion models, as presented in this paper, marks an advancement in the field of generative artificial intelligence from a privacy preservation perspective. This approach not only addresses the paramount concerns of user privacy and data confidentiality but also opens up new horizons for the application of generative models in sensitive domains such as healthcare, personal security, and confidential design, where privacy concerns have previously been a barrier.

By focusing on the optimization of the denoising step and implementing efficient computational strategies such as the distortion-based method and COO sparse tensor representations, this work mitigates the computational overhead associated with HE. This makes the proposed framework not just a theoretical contribution but a practical tool for enhancing the privacy and security of generative AI applications.

Future work could extend this framework to state-of-the-art stable diffusion implementation to support more pre-trained models. Additionally,  secure model forward needs more investigation, since it is much more computational intensive and harder for encryption. 
Cooperation between users and servers is effective for privacy preservation; however, it may lead to inefficiencies due to the heavy communication requirements.

In conclusion, this paper not only addresses a critical gap in the literature by providing a viable solution to privacy concerns in GenAI but it also sets a new standard for the integration of security considerations into the development and deployment of AI technologies. As the field continues to evolve, the principles and methodologies introduced here are likely to serve as a foundation for a new era of privacy-aware artificial intelligence, fostering innovation while safeguarding user privacy and data integrity.

\newpage





\bibliographystyle{plain}
\bibliography{main}
\end{document}